\documentclass[11pt]{article}
\usepackage{amssymb,amsmath,amsfonts}
\usepackage{graphicx}
\usepackage{graphics}
\usepackage{eepic,epsfig}

\begin{document}

\title{Dynamical Casimir effect on surface waves}
\author{A. A. Saharian\thanks{%
E-mail: saharian@ysu.am } \\
\textit{Institute of Applied Problems of Physics} \\
\textit{25 Nersessian Street, 0014 Yerevan, Armenia}}
\maketitle

\begin{abstract}
We consider the quantum radiation of scalar particles from a surface wave
excited on a plane surface of a mirror. It is assumed that the field obeys
Dirichlet condition on the boundary of the mirror. In both cases of running
and standing surface waves the expression is given for the spectral-angular
distribution of the number of the radiated quanta.
\end{abstract}

\bigskip

\begin{center}
{\small Talk presented at the International Conference \\[0pt]
"Electrons, Positrons, Neutrons and X-rays Scattering Under External
Influences"\\[0pt]
Yerevan-Meghri, Armenia, October 26-30, 2009 }
\end{center}

\bigskip

\section{Introduction}

The Casimir effect \cite{Casimir} is one of the most interesting
manifestations of nontrivial properties of the vacuum state in a quantum
field theory and can be viewed as a polarization of vacuum by boundary
conditions. A new phenomenon, a quantum creation of particles (the dynamical
Casimir effect) occurs when the geometry of the system varies in time (for a
recent review see \cite{Dodo10}). In the present talk, based on \cite%
{Grig97a,Grig97b}, we discuss the quantum radiation from a surface wave
excited on a flat boundary.

\section{General formula for the number of the radiated quanta}

\label{sec: Gen}

Consider a quantum scalar field $\varphi (x)$ in a spacetime region with the
boundary $S$. We assume that on the boundary the field obeys Dirichlet
boundary condition. The field equation and the boundary condition have the
form%
\begin{equation}
\left( \nabla _{l}\nabla ^{l}+m^{2}\right) \varphi (x)=0,\;\varphi
(x)|_{x\in S}=0.  \label{FieldEq}
\end{equation}%
In the case of a moving boundary, the interaction with quantum fluctuations
of the field can lead to the creation of real quanta out of vacuum. Let the
vector $\xi ^{l}(x)=n^{l}\xi (x)$ describes the displacement of the
hypersurface $S$ from the static hypersurface $S_{0}$: $x^{l}+\xi ^{l}(x)\in
S$ if $x^{l}\in S_{0}$. Here $n^{l}$ is the unit normal to the boundary $%
S_{0}$. Assuming that the displacement is small, in the first approximation,
for the number of quanta radiated in the interval of quantum numbers $(\nu
,\nu +d\nu )$ one has the formula \cite{Grig97a}
\begin{equation}
n(\nu )d\nu =\int d\nu ^{\prime }\left\vert \int_{S_{0}}d\Sigma
\,n^{i}n^{l}\xi (x)[\partial _{i}\varphi _{0\nu ^{\prime }}(x)][\partial
_{l}\varphi _{0\nu }(x)]\right\vert ^{2}d\nu ,  \label{nnu}
\end{equation}%
where $\{\varphi _{0\nu }(x),\varphi _{0\nu }^{\ast }(x)\}$ is a complete
set of eigenfunctions with the set of quantum numbers $\nu $ which are
solutions of the boundary-value problem (\ref{FieldEq}) with $S=S_{0}$.
Below we consider applications of the general formula (\ref{nnu}) to special
cases.

\section{Two-dimensional spacetime}

\label{sec:2Dim}

First we consider the case of a two-dimensional spacetime with the
coordinates $(t,x)$ and with the line $x=0$ as $S_{0}$. The corresponding
eigenfunctions have the form%
\begin{equation}
\varphi _{0k}(t,x)=\frac{\sin (kx)}{\sqrt{\pi \omega }}e^{-i\omega
t},\;\omega =\sqrt{k^{2}+m^{2}},  \label{phi0k}
\end{equation}%
with $0\leqslant k<\infty $. From (\ref{nnu}), for the density of the number
of quanta emitted to the region $x>0$ one finds%
\begin{equation}
n(k)=\frac{k^{2}}{\pi \omega }\int_{0}^{\infty }dk^{\prime }\frac{k^{\prime
2}}{\omega ^{\prime }}|\xi (\omega +\omega ^{\prime })|^{2},\;\omega
^{\prime }=\sqrt{k^{\prime 2}+m^{2}},  \label{nk}
\end{equation}%
where
\begin{equation}
\xi (\omega )=\int_{-\infty }^{+\infty }dt\,\xi (t)e^{-i\omega t}.
\label{ksiom}
\end{equation}%
\qquad

\begin{figure}[tbph]
\begin{center}
\epsfig{figure=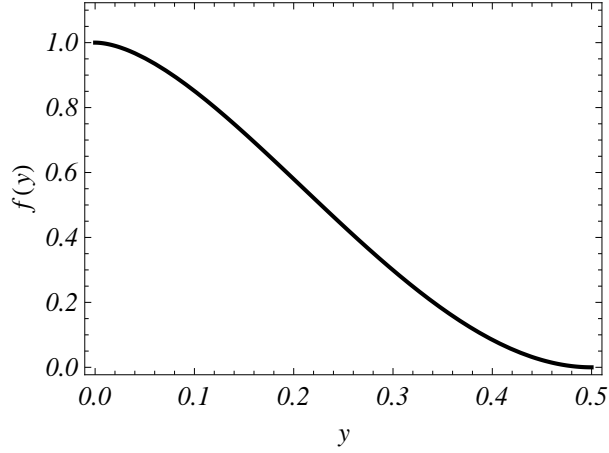,width=8.cm,height=6.cm}
\end{center}
\caption{The graph of the function $f(y)$ defined by (\protect\ref{fy}).}
\label{fig1}
\end{figure}

For harmonic oscillations of the boundary, $\xi (t)=\xi _{0}\cos (\omega
_{0}t)$, the expression for the density of the number of quanta radiated per
unit time takes the form%
\begin{equation}
n(k)=\frac{\xi _{0}^{2}k^{2}}{2\pi \omega }\sqrt{(\omega _{0}-\omega
)^{2}-m^{2}},\;m\leqslant \omega \leqslant \omega _{0}-m.  \label{nk1}
\end{equation}%
As a necessary condition for the radiation one has $\omega _{0}>2m$ and the
maximal energy of the radiated quanta is equal to $\omega _{0}-m$. The
energy radiated per unit time, evaluated by using Eq. (\ref{nk1}), is given
by%
\begin{equation}
E=\int_{0}^{\infty }dk\,n(k)\omega =\frac{\xi _{0}^{2}\omega _{0}^{4}}{24\pi
}f(m/\omega _{0}),  \label{E}
\end{equation}%
with the function%
\begin{equation}
f(y)=12\int_{y}^{1/2}dz\,\sqrt{z^{2}-y^{2}}\sqrt{(1-z)^{2}-y^{2}}%
,\;0\leqslant y\leqslant 1/2.  \label{fy}
\end{equation}%
This function is plotted in figure \ref{fig1}. From the graph it is seen
that the maximum energy is radiated in the case of a massless field.
Formulae (\ref{nk1}) and (\ref{E}) are valid under the condition $|d\xi
/dt|\sim \xi _{0}\omega _{0}\ll 1$.

For a massless field we can obtain general formula for the radiated power in
the case of general motion of the mirror. Indeed, by taking into account (%
\ref{nk}), the energy radiated per unit time to the region $x>0$ is
presented in the form%
\begin{equation}
E=\int_{0}^{\infty }d\omega \,n(\omega )\omega =\frac{2i}{\pi ^{2}}%
\int_{-\infty }^{+\infty }dt\int_{-\infty }^{+\infty }dt^{\prime }\frac{\xi
(t)\xi (t^{\prime })}{(t-t^{\prime }+i\varepsilon )^{5}},\;\varepsilon >0.
\label{E1}
\end{equation}%
Introducing new integration variables $t_{1}=t^{\prime }-t$, $%
t_{2}=t^{\prime }+t$, the integral over $t_{1}$ is evaluated by using the
relation%
\begin{equation}
\int_{-\infty }^{+\infty }dt\frac{g(t_{1})}{(t_{1}-i\varepsilon )^{5}}=\frac{%
i\pi }{4}g^{(4)}(0),  \label{rel1}
\end{equation}%
with
\begin{equation}
g(t_{1})=\xi \left( \frac{t_{2}+t_{1}}{2}\right) \xi \left( \frac{t_{2}-t_{1}%
}{2}\right) .  \label{gt1}
\end{equation}%
Integrating by parts, one finds%
\begin{equation}
E=\frac{1}{12\pi }\int_{-\infty }^{+\infty }dt\,|\xi ^{\prime \prime
}(t)|^{2}.  \label{E3}
\end{equation}%
This result coincides with the formula derived in \cite{Ford82} by using the
expectation value of the energy-momentum tensor: $E=\int dV\,\left\langle
T^{00}\right\rangle $.

\section{Examples for 4-dimensional spacetime}

\label{sec:4Dim}

In a 4-dimensional spacetime, as the hypersurface $S_{0}$ we take the plane $%
x=0$. For the corresponding eigenfunctions one has%
\begin{equation}
\varphi _{0\mathbf{k}}(t,\mathbf{x})=\frac{\sin (k_{1}x)}{2\pi \sqrt{\pi
\omega }}e^{i\mathbf{k}_{\perp }\mathbf{x}_{\perp }-i\omega t},\;\omega =%
\sqrt{k^{2}+m^{2}},  \label{phi0k1}
\end{equation}%
where $\mathbf{x}=(x,y,z)$, $\mathbf{x}_{\perp }=(y,z)$, $\mathbf{k}_{\perp
}=(k_{2},k_{3})$. For the density of the number of quanta from (\ref{nnu})
we find the expression%
\begin{equation}
n(\mathbf{k})=\frac{1}{16\pi ^{6}}\int d\mathbf{k}^{\prime }\frac{%
k_{1}^{2}k_{1}^{\prime 2}}{\omega \omega ^{\prime }}|\xi (\omega +\omega
^{\prime },\mathbf{k}_{\perp }+\mathbf{k}_{\perp }^{\prime })|^{2},
\label{nkvec}
\end{equation}%
where%
\begin{equation}
\xi (\omega ,\mathbf{k}_{\perp })=\int_{-\infty }^{+\infty }dtdydz\,\xi
(t,y,z)e^{i\mathbf{k}_{\perp }\mathbf{x}_{\perp }-i\omega t}.
\label{ksiomka}
\end{equation}

First let us consider the case when the plane oscillates as a whole: $\xi
=\xi (t)$. For a massless field we find%
\begin{equation}
n(\mathbf{k})=\frac{k_{1}^{2}}{4\pi ^{4}\omega }\int_{k_{\perp }}^{\infty
}d\omega ^{\prime }\sqrt{\omega ^{\prime 2}-k_{\perp }^{2}}|\xi (\omega
+\omega ^{\prime })|^{2}.  \label{nkvec1}
\end{equation}%
For the total energy radiated into the region $x>0$ we obtain the expression%
\begin{equation}
E=\int d\mathbf{k}\,n(\mathbf{k})\omega =\frac{1}{i\pi ^{3}}\int_{-\infty
}^{+\infty }dt\int_{-\infty }^{+\infty }dt^{\prime }\frac{\xi (t)\xi
(t^{\prime })}{(t-t^{\prime }+i\varepsilon )^{7}}.  \label{E4}
\end{equation}%
After the evaluation of the integral, in a way similar to the
two-dimensional case, we find%
\begin{equation}
E=\frac{1}{720\pi ^{2}}\int_{-\infty }^{+\infty }dt\,|\xi ^{\prime \prime
\prime }(t)|^{2}.  \label{E5}
\end{equation}%
By taking into account the radiation into the region $x<0$, this result
coincides with the formula derived in \cite{Ford82} by using the expectation
value of the energy-momentum tensor.

\begin{figure}[tbph]
\begin{center}
\epsfig{figure=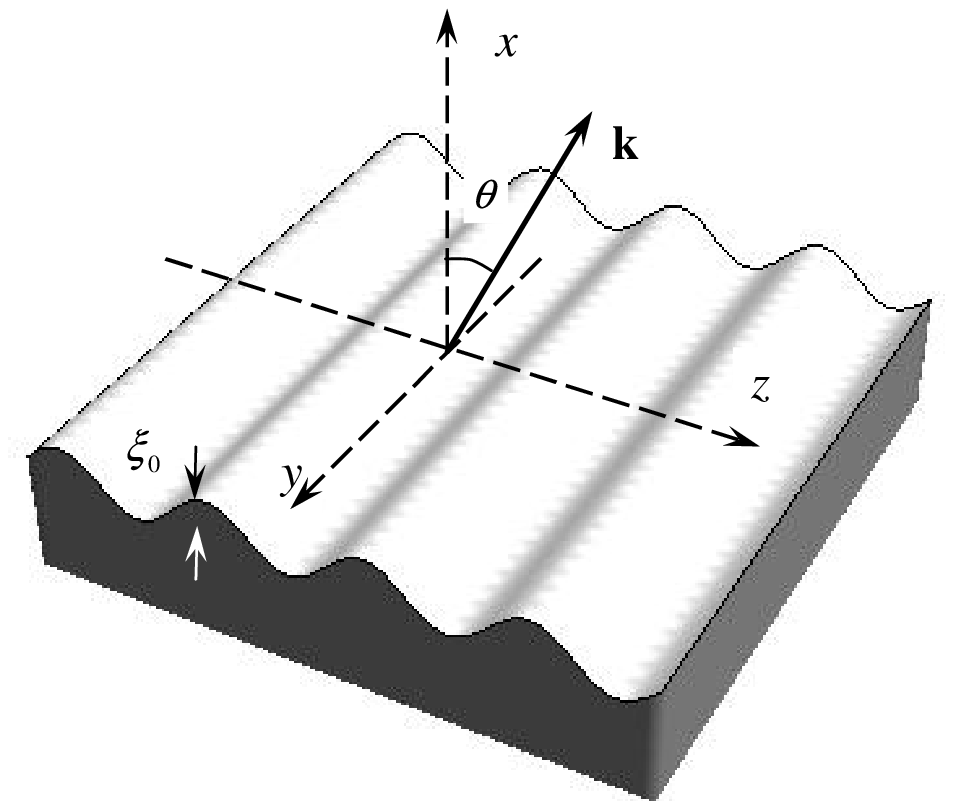,width=8.cm,height=6.cm}
\end{center}
\caption{Quantum radiation from a surface wave excited on a plane boundary.}
\label{fig2}
\end{figure}

Now we turn to the case of a running surface wave excited on the surface of
a plane mirror (see figure \ref{fig2}). The corresponding displacement
function has the form%
\begin{equation}
\xi (t,y,z)=f(z-v_{0}t).  \label{ksiSW1}
\end{equation}%
In this case for the Fourier-transform we have%
\begin{equation}
\xi (\omega ,\mathbf{k}_{\perp })=(2\pi )^{2}\delta (k_{2})\delta (\omega
-k_{3}v_{0})\int_{-\infty }^{+\infty }dz\,f(z)e^{ik_{3}z}.  \label{ksiSW2}
\end{equation}%
One has $k_{3}\leqslant \omega $ and for $v_{0}<1$ this expression vanishes.
In this case the radiation is absent. This result becomes obvious if we pass
to the reference frame moving with the surface wave. In the case of harmonic
oscillations,%
\begin{equation}
\xi (t,y,z)=\xi _{0}\cos (k_{0}z-\omega _{0}t),  \label{ksiSWharm}
\end{equation}%
with $\omega _{0}>k_{0}$, the quanta are radiated satisfying the condition%
\begin{equation}
\omega _{0}-\omega \geqslant \sqrt{k_{2}^{2}+(k_{0}-k_{3})^{2}}.  \label{om0}
\end{equation}%
The number density of the quanta radiated from unit surface of the mirror
per unit time is given by the expression%
\begin{equation}
n(\mathbf{k})=\frac{k_{1}^{2}\xi _{0}^{2}}{8\pi ^{3}\omega }[(\omega
_{0}-\omega )^{2}-k_{2}^{2}-(k_{0}-k_{3})^{2}]^{1/2}.  \label{nkvec2}
\end{equation}%
Introducing spherical coordinates in the momentum space, $k_{1}=\omega \cos
\theta $, $k_{2}=\omega \sin \theta \sin \phi $, $k_{3}=\omega \sin \theta
\cos \phi $, $0\leqslant \theta \leqslant \pi /2$, $0\leqslant \phi
\leqslant 2\pi $, for the spectral-angular density of the number of the
radiated quanta we have
\begin{equation}
\frac{dN}{d\omega d\Omega }=\frac{\xi _{0}^{2}\omega ^{3}}{8\pi ^{3}}\cos
^{2}\theta (\omega _{0}^{2}-2\omega _{0}\omega -k_{0}^{2}+2k_{0}\omega \sin
\theta \cos \phi +\omega ^{2}\cos ^{2}\theta )^{1/2},  \label{dN}
\end{equation}%
where $d\Omega =\sin \theta d\theta d\phi $ is the solid angle element and $%
\theta $ is the angle between the normal to the mirror and the direction of
the radiation (see figure \ref{fig2}). For a given $\omega $, the allowed
angular region for the radiation is determined by the non-negativity
condition for the expression under the square root in (\ref{dN}).

\begin{figure}[tbph]
\begin{center}
\epsfig{figure=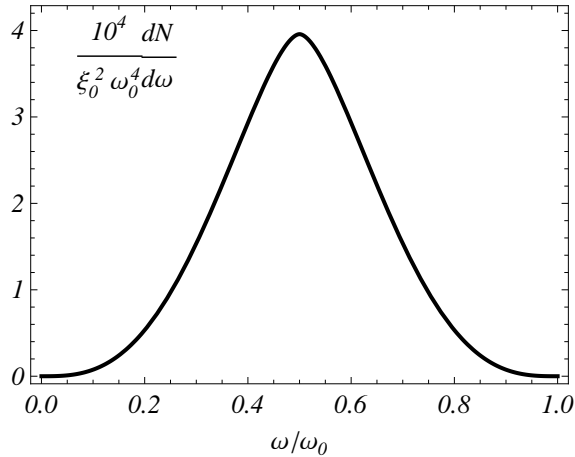,width=8.cm,height=6.cm}
\end{center}
\caption{The spectral density of the radiated quanta given by (\protect\ref%
{nom}).}
\label{fig3}
\end{figure}

The limit $k_{0}\rightarrow 0$ of this formula corresponds to the mirror
oscillating as a whole. In this case, for the number of quanta radiated in
the energy range $(\omega ,\omega +d\omega )$, in the solid angle $d\Omega $
from the unit surface of the mirror per unit time one has%
\begin{eqnarray}
\frac{dN}{d\omega d\Omega } &=&\frac{\xi _{0}^{2}\omega ^{3}}{8\pi ^{3}}\cos
^{2}\theta (\omega _{0}^{2}-2\omega _{0}\omega +\omega ^{2}\cos ^{2}\theta
)^{1/2},  \notag \\
\sin \theta  &\leqslant &\omega _{0}/\omega -1,\;0\leqslant \theta \leqslant
\pi /2.  \label{nkvec3}
\end{eqnarray}%
The spectral density of the radiation is obtained after the integration of (%
\ref{nkvec3}) over the solid angle: $dN/d\omega =\omega ^{2}\int n(\mathbf{k}%
)d\Omega $. This leads to the result
\begin{equation}
\frac{dN}{d\omega }=\frac{\xi _{0}^{2}\omega _{0}^{4}}{32\pi ^{2}}\left[
u(1-u)^{3}+u^{3}(1-u)+(1/2)(1-2u)^{2}\ln |1-2u|\right] ,  \label{nom}
\end{equation}%
with the notation $u=\omega /\omega _{0}$. The expression on the right-hand
side of (\ref{nom}) takes its maximum value at $u=1/2$ with $(dN/d\omega
)_{\max }=\xi _{0}^{2}\omega _{0}^{4}/(2^{8}\pi ^{2})$. In figure \ref{fig3}
we have plotted the spectral density of the number of the radiated quanta as
a function of $\omega /\omega _{0}$.

For the total number of the radiated quanta and the total energy (per unit
surface and per unit time) one finds (in standard units, for the radiated
energy see also \cite{Ford82})%
\begin{equation}
N=\frac{\xi _{0}^{2}\omega _{0}^{5}}{720\pi ^{2}c^{4}},\;E=\frac{\hbar \xi
_{0}^{2}\omega _{0}^{6}}{1440\pi ^{2}c^{4}}.  \label{NE}
\end{equation}%
The Planck constant in the expression for the radiated energy shows the
quantum nature of the radiation.

Now we turn to the case of a standing surface wave excited on the surface of
a plane mirror in the strip $0\leqslant z\leqslant l$, $-\infty <y<+\infty $:%
\begin{equation}
\xi (t,y,z)=\xi _{0}\cos (\omega _{0}t)\sin (k_{0}z),\;k_{0}=\pi
n/l,\;n=1,2,3,\ldots ,  \label{ksiSt1}
\end{equation}%
for $0\leqslant z\leqslant l$ and $\xi (t,y,z)=0$ otherwise. For the number
of the radiated quanta per unit time and per unit length along $y$ we find%
\begin{equation}
n(\mathbf{k})=\frac{k_{1}^{2}\xi _{0}^{2}}{8\pi ^{4}\omega }%
\int_{w_{1}}^{w_{2}}du\,\frac{\sqrt{(w-w_{1})(w_{2}-w)}}{(1-w^{2})^{2}}\left[
1-(-1)^{n}\cos (\pi nw)\right] ,  \label{nkvec4}
\end{equation}%
where%
\begin{equation}
w_{2,1}=[k_{3}\pm \sqrt{(\omega _{0}-\omega )^{2}-k_{2}^{2}}]/k_{0}.
\label{u21}
\end{equation}%
The following conditions should be satisfied%
\begin{equation}
\omega =\sqrt{k_{1}^{2}+k_{2}^{2}+k_{3}^{2}}\leqslant \omega
_{0},\;|k_{2}|\leqslant \omega _{0}-\omega ,  \label{Conds}
\end{equation}%
for the presence of the radiation. In particular, the second inequality
imposes a constraint on the angular region for the radiation. Note that the
integrand in (\ref{nkvec4}) is finite at $u=\pm 1$.

\section{Conclusion}

\label{sec:Conc}

In the present talk we have discussed the quantum radiation from surface
waves excited on a plane boundary. For simplicity we have considered the
case of a scalar field with Dirichlet boundary condition. Other boundary
conditions can be treated in a similar way. Assuming that the displacement
of the moving boundary from the static one is small, in the first
approximation, the expression for the density of the number of radiated
quanta has the form (\ref{nnu}). We give applications of this general
formula to various special cases. First, we have considered the simplest
case of two-dimensional spacetime when the general formula takes the form (%
\ref{nk}). For harmonic oscillations of the boundary this formula simplifies
to (\ref{nk1}). As a necessary condition for the presence of the quantum
radiation one has $\omega _{0}>2m$ and the maximal energy of the radiated
quanta is equal to $\omega _{0}-m$. On the base of the number of the
radiated quanta, we have evaluated the radiated energy. The latter coincides
with the corresponding result evaluated by using the renormalized
expectation value of the energy-momentum tensor.

More realistic case of 4-dimensional spacetime is discussed in
section \ref{sec:4Dim} with the expression (\ref{nkvec}) for the
number of the radiated quanta. For a massless field we have
considered various special cases. When the plane boundary
oscillates as a whole the number of the radiated quanta and the
total radiated energy are given by expressions (\ref{nkvec1}) and (\ref{E5}%
), respectively. As in the two-dimensional case, the latter coincides with
the result obtained on the base of the expectation value of the
energy-momentum tensor. Next, we have considered the cases of running and
standing surface waves. For the running wave the radiation is present in the
case when the phase velocity of the wave is larger than the speed of light.
The spectral-angular distribution of the radiated quanta is given by
expression (\ref{nkvec3}) for a running wave and by (\ref{nkvec4}) for a
standing surface wave.

The author gratefully acknowledges the organizers of the International
Conference "Electrons, Positrons, Neutrons and X-rays Scattering Under
External Influences", Yerevan-Meghri, October 26-30, 2009, for the financial
support to attend the conference.


\begin{thebibliography}{9}
\bibitem{Casimir} V.M. Mostepanenko, N.N. Trunov, The Casimir Effect and Its
Applications (Clarendon, Oxford, 1997); K.A. Milton, The Casimir Effect:
Physical Manifestation of Zero-Point Energy (World Scientific, Singapore,
2002); V.A. Parsegian, Van der Waals Forces (Cambridge University Press,
Cambridge, 2005); M. Bordag, G.L. Klimchitskaya, U. Mohideen, and V.M.
Mostepanenko, Advances in the Casimir Effect (Oxford University Press,
Oxford, 2009).

\bibitem{Grig97a} L.Sh. Grigoryan, A.A. Saharian, Izvestia NAN Armenii,
Fizika \textbf{32}, 223 (1997) [English translation J. Contemp. Phys.].

\bibitem{Grig97b} L.Sh. Grigoryan, A.A. Saharian, Izvestia NAN Armenii,
Fizika \textbf{32}, 275 (1997) [English translation J. Contemp. Phys.].

\bibitem{Dodo10} V.V. Dodonov, arXiv:1004.3301 (2010); D.A.R. Dalvit, P.A.
Maia Neto, F.D. Mazzitelli, arXiv:1006.4790 (2010).

\bibitem{Ford82} L.H. Ford, A. Vilenkin, Phys. Rev. D \textbf{25}, 2569
(1982).
\end{thebibliography}
\end{document}